\documentclass[conference]{IEEEtran}
\IEEEoverridecommandlockouts
\usepackage{cite}
\usepackage{amsmath,amssymb,amsfonts}
\usepackage{algorithmic}
\usepackage{graphicx}
\usepackage{textcomp}
\usepackage{xcolor}
\usepackage{algorithm}
\usepackage{amsfonts}
\usepackage{cite}
\usepackage{graphicx}
\usepackage{amssymb}
\usepackage{caption}
\usepackage{epsfig}
\usepackage{subfigure}
\usepackage{amsmath}
\usepackage{epstopdf}

\usepackage{array}
\usepackage{color}
\usepackage{amsmath}
\usepackage{longtable}
\usepackage{algorithmic}
\UseRawInputEncoding
\usepackage{algorithm}
\usepackage{cases}
\usepackage{mathrsfs}
\usepackage{psfrag}
\usepackage{url}
\usepackage{setspace}

\usepackage{etoolbox}
\makeatletter
\patchcmd{\@makecaption}
  {\scshape}
  {}
  {}
  {}
\makeatletter
\patchcmd{\@makecaption}
  {\\}
  {.\ }
  {}
  {}
\makeatother

\newtheorem{Thm}{Theorem}

\newtheorem{Prob}{Problem}

\newcounter{TempEqCnt}

\IEEEoverridecommandlockouts

\newcommand{\blue}[1]{\textcolor{black}{#1}}

\def\BibTeX{{\rm B\kern-.05em{\sc i\kern-.025em b}\kern-.08em
    T\kern-.1667em\lower.7ex\hbox{E}\kern-.125emX}}
\begin{document}

\title{Device Activity Detection for Massive Grant-Free Access Under Frequency-Selective Rayleigh Fading\\
}

\author{\IEEEauthorblockN{Yuhang Jia}
\IEEEauthorblockA{Shanghai Jiao Tong Univ., China}
\IEEEauthorblockA{Jay\_Yoga@sjtu.edu.cn}
\and
\IEEEauthorblockN{Ying Cui}
\IEEEauthorblockA{Shanghai Jiao Tong Univ., China}
\IEEEauthorblockA{cuiying@sjtu.edu.cn}
\and
\IEEEauthorblockN{Wuyang Jiang}
\IEEEauthorblockA{Shanghai Univ. of Engineering Science, China}
\IEEEauthorblockA{jiang-wuyang@sues.edu.cn}
\thanks{This work was supported in part by the National Key Research and Development Program of China under Grant 2018YFB1801102 and in part by the Natural Science Foundation of Shanghai under Grant 20ZR1425300.
}
}


\maketitle

\begin{abstract}
Device activity detection and channel estimation for massive grant-free  access under frequency-selective fading have unfortunately been an outstanding problem. This paper aims to address the challenge. Specifically, we present an orthogonal frequency division multiplexing (OFDM)-based massive grant-free  access scheme for a wideband system with one $M$-antenna base station (BS), $N$ single-antenna Internet of Things (IoT) devices, and $P$ channel taps. We obtain two different but equivalent models for the received pilot signals under frequency-selective Rayleigh fading. Based on each model, we formulate device activity detection as a non-convex maximum likelihood estimation (MLE) problem and propose an iterative algorithm to obtain a stationary point using optimal techniques. The two proposed MLE-based methods have the identical computational complexity order $\mathcal{O}(NPL^2)$, irrespective of $M$, and degrade to the existing MLE-based device activity detection method when $P=1$. Conventional channel estimation methods can be readily applied for channel estimation of detected active devices under frequency-selective Rayleigh fading, based on one of the derived models for the received pilot signals. Numerical results show that the two proposed methods have different preferable system parameters and complement each other to offer promising device activity detection design for grant-free massive access under frequency-selective Rayleigh fading.
\end{abstract}
\section{Introduction}
Driven by the proliferation of the Internet of Things (IoT), massive machine-type communication (mMTC) plays a vital role in the fifth generation (5G) cellular technologies and beyond. It is incredibly challenging to support enormous IoT devices which are energy-limited and sporadically active, and have little data to send once activate. Massive grant-free access with multiple-input multiple-output (MIMO) has been recently proposed to address the challenge. Specifically, devices are pre-assigned specific non-orthogonal pilots, active devices directly send their pilots, and the base station (BS) detects the active devices and estimates their channel conditions from the received signal of non-orthogonal pilots \cite{Erik}. Unfortunately, vast potential non-orthogonal pilots complicate the signal processing at the BS.

Due to inherent sparse device activities in massive grant-free access, joint device activity detection and channel estimation can be formulated as compressed sensing (CS) problems and solved by CS-based algorithms \cite{Liu18TSP, grouplasso, JSAC_li}. Specifically, \cite{Liu18TSP} proposes an approximate message passing (AMP) algorithm with a minimum mean square error (MMSE) estimation denoiser. In \cite{JSAC_li}, the authors propose an alternating direction method of multipliers (ADMM)-based algorithm for GROUP LASSO \cite{grouplasso}. Besides, several works focus only on device activity detection, as conventional channel estimation methods can be directly applied for estimating channel conditions of the detected active devices. For instance, \cite{Caire18ISIT} formulates device activity detection as a maximum likelihood estimation (MLE) problem and proposes a coordinate descent method to obtain a stationary point. This MLE-based method is also analyzed in \cite{Caire18ISIT,Yu19ICC}. Motivated by \cite{Caire18ISIT}, \cite{Jiang21TWC} formulates device activity detection with prior activity distribution as a maximum a posteriori probability (MAP) estimation problem and extends the coordinate descent method in \cite{Caire18ISIT} to obtain a stationary point.

It is worth noting that all existing works \cite{Liu18TSP,JSAC_li,Caire18ISIT,Yu19ICC,Jiang21TWC} consider massive grant-free access for a narrow band system under flat fading.  However, due to the signal corruption under frequency-selective fading, the existing methods for activity detection and channel estimation designed for a narrow
band system under flat fading are no longer applicable for a wideband system under frequency-selective fading.  On the other hand, orthogonal frequency division multiplexing (OFDM) provides a high degree of robustness against channel-frequency selectivity. It hence is an attractive choice for 4G-LTE and 5G-NR. In this paper, we would like to shed some light toward this direction. Specifically, we present an orthogonal frequency division multiplexing (OFDM)-based massive grant-free access scheme with one $M$-antenna BS, $N$ single-antenna IoT devices, and $P$ channel taps. We obtain two different but equivalent models for the received pilot signals under frequency-selective Rayleigh fading. Based on each model, we formulate device activity detection as a non-convex MLE problem and propose an iterative algorithm to obtain a stationary point using optimal techniques. The two proposed MLE-based device activity detection methods have the identical computational complexity order $\mathcal{O}(NPL^2)$ and degrade to the existing MLE-based device activity detection method \cite{Caire18ISIT,Yu19ICC} when $P=1$. Notice that conventional channel estimation methods can be readily applied for channel estimation of detected active devices under frequency-selective Rayleigh fading, based on one of the derived models for the received pilot signals. Numerical results show that the two proposed methods offer promising device activity detection design for frequency-selective Rayleigh fading. Furthermore, one method always achieves a lower error rate than the other with a shorter computation time if $P$ is small and a longer computation time otherwise. This is the first work investigating massive grant-free access under frequency-selective fading to the best of our knowledge.

$\mathbf{Notation:}$ We represent vectors by boldface lowercase letters (e.g., $\mathbf{x}$), matrices by boldface uppercase letters (e.g., $\mathbf{X}$), scalar constants by non-boldface letters (e.g., $x$), and sets by calligraphic letters (e.g., $\mathcal{X}$).
The notation $x_i$ represents the $i$-th element of vector $\mathbf{x}$, $\mathbf{X}_{i,:}$ represents the $i$-th row of matrix $\mathbf{X}$, and $\mathbf{X}_{:,i}$ represents the $i$-th column of matrix $\mathbf{X}$. $\mathbf X_{:,1:K}$ represents the matrix consisting of the first $K$ columns of the matrix $\mathbf{X}$. $\mathbf{X}^H$ and $\text{tr}\left(\mathbf{X}\right)$ denote the conjugate transpose and trace of the matrix $\mathbf{X}$, respectively.
$\rm diag \left(\mathbf{x}\right)$ is a diagonal matrix with the entries of $\mathbf{x}$ on its main diagonal. $\lvert \cdot \rvert$ denotes the modulus of a complex number. The complex field and real field are denoted by $\mathbb{C}$ and $\mathbb{R}$, respectively. $\otimes$ denotes the Kronecker product. $\mathbf{I}_L$ and $\mathbf{e}_n$ denotes the $L\times L$ identity matrix and a unit vector whose $n$-th component is $1$, all others $0$. $\text{Pr}[x]$ denotes the probability of the event $x$.
\setcounter{TempEqCnt}{\value{equation}}
\setcounter{equation}{10}
\begin{figure*}
\begin{align}
  f^{(1)}_{\boldsymbol\alpha,n}(d) \triangleq &
 \log |\mathbf{I}_P+dg_n\mathbf S_{n}^H\mathbf \Sigma^{(1)-1}_{\boldsymbol\alpha}\mathbf S_{n}| + dg_n\text{tr}\left((\mathbf{I}_P+dg_n\mathbf{S}_n^H\boldsymbol\Sigma_{\boldsymbol\alpha}^{(1)-1}
\mathbf{S}_n)^{-1}\mathbf{S}_n^H\boldsymbol\Sigma_{\boldsymbol\alpha}^{(1)-1}
\widehat{\mathbf \Sigma}_{\mathbf R}\boldsymbol\Sigma_{\boldsymbol\alpha}^{(1)-1}\mathbf{S}_n\right)\label{eq:function} \\
g^{(1)}_{\boldsymbol\alpha,n}(d) \triangleq &
d^{2P-1} \sum_{p\in\mathcal{P}}v_p^2h(\mathbf{v}_{-p},2P-1) +
\sum_{t=0}^{2P-2}d^t\sum_{p\in\mathcal{P}}
(v_p^2+v_p-u_p)h(\mathbf{v}_{-p},t)\label{eq:derivative_function}
\end{align}
\hrulefill
\vspace{-1mm}
\end{figure*}
\setcounter{equation}{\value{TempEqCnt}}
\section{System Model}\label{sec:system}
We consider a single-cell cellular network with one $M$-antenna BS and $N$ single-antenna IoT devices. Let $\mathcal M\triangleq \{1,2,\cdots,M\}$ and $\mathcal N\triangleq \{1,2,\cdots,N\}$ denote the sets of device and antenna indices, respectively. \blue{For all $n\in\mathcal{N}$, l}et $g_n>0$ denote the large-scale fading power of the channel between device $n$ and the BS. Small-scale fading follows the block fading model, i.e., small-scale fading coefficients remain constant within each coherence block and are \blue{independent and identically distributed (i.i.d.)} over coherence blocks. We consider a wideband system and adopt the frequency-selective Rayleigh fading channel model for small-scale fading. Let $P$ denote the number of channel taps, and let $\mathcal P\triangleq \{1,2,\cdots,P\}$ denote the set of channel tap indices. Denote $h_{n,m,p} \in \mathbb{C}$ as the $p$-th coefficient of the channel impulse response (CIR) of the channel between device $n$ and the BS over antenna $m$, for all $n\in\mathcal{N},m\in\mathcal{M}$, $p\in\mathcal{P}$. We assume $h_{n,m,p}\sim C\mathcal{N}(0,1), n\in\mathcal{N}, m \in \mathcal{M}, p\in\mathcal{P}$.

We study the massive access scenario arising from mMTC, where very few devices among a large number of potential devices are active and access the BS in each coherence block. For all $n\in\mathcal{N}$, let $\alpha_n\in\{0,1\}$ denote the activity state of device $n$, where $\alpha_n=1$ indicates that device $n$ is active and $\alpha_n=0$ otherwise. In the considered massive access scenario, $\sum_{n\in\mathcal{N}} \alpha_n \ll N$, i.e., $\boldsymbol{\alpha}\triangleq (\alpha_n)_{n\in\mathcal N} \in \{0,1\}^N$ is sparse. We adopt an OFDM-based massive grant-free access scheme. Let $L$ denote the number of subcarriers, and denote $\mathcal L\triangleq \{1,2,\cdots,L\}$ as the set of subcarrier indices. Assume $P<L$. Each device $n\in\mathcal{N}$ is pre-assigned a specific pilot sequence $\tilde{\mathbf s}_n\triangleq(\tilde{s}_{n,\ell})_{\ell\in\mathcal L}\in \mathbb C^{L}$ consisting of $L\ll N$ OFDM symbols, each carried by one subcarrier. In the pilot transmission phase, active devices simultaneously send their length-$L$ pilots to the BS over the $L$ subcarriers, and the BS detects the activity  states of all devices and estimates the channel states of all active devices from the \blue{$LM$} received OFDM symbols over the $M$ antennas. In this paper, we focus on device activity detection under frequency-selective Rayleigh fading, which is more challenging than device activity detection under flat Rayleigh fading \cite{Erik,Liu18TSP,JSAC_li,Caire18ISIT,Yu19ICC,Jiang21TWC}. We shall see that based on one of the derived models for the received pilot signals, conventional channel estimation methods can be readily applied for channel estimation of detected active devices.

The time domain representation of the OFDM symbols in $\tilde{\mathbf{s}}_n\in\mathbb{C}^L$, i.e., the normalized inverse discrete Fourier transform (IDFT) of $\tilde{\mathbf{s}}_n$, is given by:
\begin{align}
\mathbf s_n =\mathbf F^H\tilde{\mathbf s}_n \in \mathbb{C}^{L},\quad n \in \mathcal{N}.
\label{eq:IDFTzh}
\end{align}
Here, $\mathbf F \triangleq (F_{\ell,\ell'})_{\ell,\ell'\in\mathcal{L}} \in\mathbb C^{L\times L}$ denotes the discrete Fourier transform (DFT) matrix where $F_{\ell,\ell'}\triangleq \frac{1}{\sqrt{L}}e^{-\frac{j2\pi (\ell-1) (\ell'-1)}{L}}$. At each device $n\in\mathcal{N}$, a cyclic prefix is appended to $\mathbf s_n$ before transmission. After removing the signal corresponding to the cyclic prefixes,
the received signal over the $L$ signal dimensions at antenna $m\in\mathcal{M}$, denoted as $\mathbf r_m \triangleq (r_{\ell,m})_{\ell\in\mathcal L}\in\mathbb{C}^{L}$, can be written as \cite{8421267}:
\begin{align}
\mathbf r_m =& \sum_{n\in\mathcal N} \alpha_n g_n^{\frac{1}{2}}\mathbf H_{n,m}\mathbf s_n + \mathbf n_m \notag \\ = & \sum_{n\in\mathcal N} \alpha_n g_n^{\frac{1}{2}}\mathbf H_{n,m}\mathbf{F}^{H} \tilde{\mathbf s}_n + \mathbf n_m, \quad m\in\mathcal{M},
\label{eq:receivesignal}
\end{align}
where
\begin{align}
\mathbf H_{n,m}\triangleq\begin{bmatrix}
h_{n,m,1} & h_{n,m,L} &\cdots & h_{n,m,2}\\
h_{n,m,2} & h_{n,m,1} &\cdots & h_{n,m,3}\\
\vdots & \vdots &\ddots & \vdots\\
h_{n,m,L} & h_{n,m,L-1}&\cdots & h_{n,m,1}\\
\end{bmatrix} \in \mathbb{C}^{L\times L},\label{eq:matrixH}
\end{align}
and $\mathbf{n}_m \triangleq (n_{\ell,m})_{l\in\mathcal{L}} \in \mathbb{C}^{L}$ with $n_{\ell,m}\sim\mathcal {CN}(0,\sigma^2)$ is the  additive white Gaussian noise (AWGN).
Here, for notation convenience, we let $h_{n,m,p}=0$, $p \in \mathcal{L}\backslash\mathcal{P}, n\in\mathcal N$, $m\in\mathcal M$. Note that for all $n\in\mathcal{N},m\in\mathcal{M}$, each of $h_{n,m,l},l\in\mathcal{L}$ appears $L$ times in $\mathbf{H}_{n,m}$.

For tractability, we obtain an equivalent expression of $\mathbf{r}_m$ in \eqref{eq:receivesignal} in the following \cite{8421267}. Define $\tilde{\mathbf n}_m \triangleq \mathbf{F}\mathbf{n}_m\in\mathbb{C}^{L}$.
First, we obtain the received signal in the frequency domain, i.e.,
\begin{align}
\tilde{\mathbf r}_m=&\mathbf F\mathbf r_m = \sum_{n\in\mathcal N}\alpha_n g_n^{\frac{1}{2}}\mathbf F\mathbf H_{n,m}\mathbf F^H\tilde{\mathbf s}_n + \tilde{\mathbf n}_m \notag \\
=& \sum_{n\in\mathcal N}\alpha_ng_n^{\frac{1}{2}}{\rm diag}(\tilde{\mathbf{s}}_n)\mathbf{F}(\mathbf H_{n,m})_{:,1}+\tilde{\mathbf n}_m,\quad m \in \mathcal{M}, \label{eqn:received_signal_frequency}
\end{align}
where the last equality is due to the fact that $\mathbf F\mathbf H_{n,m}\mathbf F^H \in \mathbb{C}^{L\times L}$ is a diagonal matrix \cite[Lemma 1]{8421267}. Define $\mathbf{S}_n \triangleq (\mathbf F^H{\rm diag}(\tilde{\mathbf{s}}_n)\mathbf F)_{:,1:P}$ and $\mathbf h_{n,m} \triangleq (h_{n,m,p})_{p\in\mathcal{P}}$. Then,
applying normalized IDFT to $\tilde{\mathbf r}_m$ in \eqref{eqn:received_signal_frequency}, we rewrite $\mathbf{r}_m$ in \eqref{eq:receivesignal} as:
\begin{align}
\mathbf r_m = & \mathbf F^H\tilde{\mathbf r}_m=\sum_{n\in\mathcal N} \alpha_n g_n^{\frac{1}{2}}\mathbf F^H\blue{{\rm diag}(\tilde{\mathbf{s}}_n)}\mathbf F(\mathbf H_{n,m})_{:,1}+\mathbf n_m \notag \\
 = & \sum_{n\in\mathcal N} \alpha_n g_n^{\frac{1}{2}} \mathbf{S}_n \mathbf h_{n,m} + \mathbf n_m, \quad m\in\mathcal{M}, \label{eqn:received_signal_time}
\end{align}
where the last equality is due to $\mathbf{F}^H\mathbf{F}=\mathbf{I}_L$ and $h_{n,m,p}=0$, $p \in \mathcal{L}\backslash\mathcal{P}, n\in\mathcal N$, $m\in\mathcal M$.
In contrast with $\mathbf{H}_{n,m},n\in\mathcal{N},m\in\mathcal{M}$, all elements of $\mathbf{h}_{n,m}, n\in\mathcal{N},m\in\mathcal{M}$ are i.i.d. according to $\mathcal{CN}(0,1)$\blue{, making device activity detection from $\mathbf{r}_m$ in \eqref{eqn:received_signal_time} more tractable than from $\mathbf{r}_m$ in \eqref{eq:receivesignal}.}

For ease of exposition, we assume that the large-scale fading powers, $g_n,n\in\mathcal{N}$, are known to the BS and propose two MLE-based device activity detection methods in Section~\ref{sec:non_extended} and Section~\ref{sec:ML}, respectively. The proposed methods can be readily extended to device activity detection with unknown large-scale fading powers \cite{Caire18ISIT}.  Later in Section~\ref{sec:simulation}, we shall see that compared to the method in Section~\ref{sec:ML}, the method in Section~\ref{sec:non_extended} achieves high detection accuracy for all $P$, short computation time for small $P$, and long computation time for large $P$. Therefore, we can apply them according to practical system parameters and requirements.
\section{MLE-based Device Activity Detection Using Coordinate Descent Method}\label{sec:non_extended}
In this section, we propose an MLE-based device activity detection method based on the expression of $\mathbf{r}_m$ in \eqref{eqn:received_signal_time} and the coordinate descent method.
\subsection{Problem Formulation}
$\mathbf h_{n,m}, n\in\mathcal{N}, m\in\mathcal{M}$ are i.i.d. according to ${\mathcal CN}(\mathbf 0,\mathbf I_{P})$. Thus, when $\alpha_n$, $g_n,n\in\mathcal{N}$ are  given, $\mathbf r_m, m \in\mathcal{M}$, with $\mathbf{r}_m$ given by \eqref{eqn:received_signal_time}, are i.i.d. according to
$ \mathcal{CN}\left(\mathbf 0,\boldsymbol\Sigma^{(1)}_{\boldsymbol\alpha} \right)$ \blue{\cite{Caire18ISIT}}, where
\begin{align}
\boldsymbol\Sigma^{(1)}_{\boldsymbol\alpha} \triangleq \sum_{n\in\mathcal N} \alpha_n g_n \mathbf{S}_n\mathbf S_n^H+\sigma^2\mathbf I_L. \label{eq:sigma_alpha_1}
\end{align}
Note that $\boldsymbol\Sigma^{(1)}_{\boldsymbol\alpha}$ depends on $\boldsymbol\alpha$. Let $\mathbf{R}$ with $\mathbf R_{:,m}\triangleq\mathbf{r}_m, m\in\mathcal{M}$ denote the received signal over the $M$ antennas.
Thus, the likelihood function of $\mathbf  R$, viewed as a function of $\boldsymbol\alpha$, 
is given by:
\begin{align}
&p^{(1)}(\mathbf R;\boldsymbol\alpha) \triangleq \frac{\exp\left(-\text{tr}\left(
 \boldsymbol\Sigma^{(1)-1}_{\boldsymbol\alpha}\mathbf R\mathbf R^H\right)\right)}{\pi^{LM}\vert\boldsymbol\Sigma^{(1)}_{\boldsymbol\alpha}\vert^{M}}.\label{eqn:likelihood_no_coop}
\end{align}
The maximization of $p^{(1)}(\mathbf R;\boldsymbol\alpha)$ is equivalent to the minimization of $f^{(1)}(\boldsymbol\alpha)$, where
\begin{align}
f^{(1)}(\boldsymbol\alpha) \triangleq & -\log p^{(1)}(\mathbf R;\boldsymbol\alpha) -L\log\pi \nonumber \\ = & \log|\boldsymbol\Sigma^{(1)}_{\boldsymbol\alpha}|+\text{tr}\left(\boldsymbol\Sigma_{\boldsymbol\alpha}^{(1)-1}\widehat{\mathbf \Sigma}_{\mathbf R}\right).\label{eqn:f_ml}
\end{align}
Here, $\widehat{\mathbf \Sigma}_{\mathbf R}\triangleq \frac{1}{M}\mathbf{R}\mathbf{R}^{H}$ represents the sample covariance matrix of $\mathbf{r}_m,m\in\mathcal{M}$. Note that $\widehat{\mathbf \Sigma}_{\mathbf R}$ is a sufficient statistics since $f^{(1)}(\boldsymbol\alpha)$ depends on $\mathbf{R}$ only through $\widehat{\mathbf \Sigma}_{\mathbf R}$.
Thus, the MLE problem of $\boldsymbol\alpha$ can be formulated as:\footnote{\blue{In this paper,} binary condition $\alpha_n\in\{0,1\}$ is relaxed to \blue{continuous condition} $\alpha_n\in[0,1]$ in \blue{each} estimation problem, and \blue{binary detection results are obtained} by performing thresholding \blue{after solving the estimation problem as in \cite{Caire18ISIT,Yu19ICC}.}}
\begin{Prob}[\blue{MLE for Activity Detection of Actual Devices}]\label{Prob:ML_a_non_extension}
\begin{align}
\min_{\boldsymbol\alpha} &\quad f^{(1)}(\boldsymbol\alpha)\notag \\
s.t. &\quad   \alpha_{n} \in [0,1],\quad n \in\mathcal{N}. \label{eqn:a_n}
\end{align}
\end{Prob}

\blue{Problem~\ref{Prob:ML_a_non_extension} is a non-convex optimization problem. When $P=1$, Problem~\ref{Prob:ML_a_non_extension} is equivalent to the MLE problem for activity detection of $N$ devices under flat Rayleigh fading in \cite{Caire18ISIT} and can be converted to the same form as the one in \cite{Caire18ISIT}. When $P\in\{2,3,...\}$, Problem~\ref{Prob:ML_a_non_extension} is different from the one in \cite{Caire18ISIT} and cannot be converted to its form (as $\mathbf{S}_n \mathbf{S}_n^H \in\mathbb{C}^{L\times L}$ is not a rank-one matrix).  Later, we shall see that this slight difference causes a significant challenge for solving Problem~\ref{Prob:ML_a_non_extension}.}
\subsection{Solution}
The goal of solving a non-convex problem is usually to obtain a stationary point of the problem.
We adopt the coordinate descent method
to obtain a stationary point of Problem~\ref{Prob:ML_a_non_extension}.
Specifically, given $\boldsymbol\alpha$ obtained in the previous step, the coordinate descent optimization w.r.t. $\alpha_n$ is equivalent to the optimization of the increment $d$ in $\alpha_n$ \cite{Caire18ISIT}:
\begin{align}
\min_{d\in [-\alpha_n,1-\alpha_n]} \ f^{(1)}(\boldsymbol\alpha + d\mathbf e_n).\label{eqn:Penal_a_ML_extension}
\end{align}
We shall see that it is more challenging to solve the coordinate descent optimization for $P\in\{2,3,...\}$ in~\eqref{eqn:Penal_a_ML_extension} than to solve that for $P=1$. In the following, we define two important functions based on which we can characterize the optimal solution of the problem in \eqref{eqn:Penal_a_ML_extension}. Specifically, we first define $f^{(1)}_{\boldsymbol\alpha,n}(d)$ in \eqref{eq:function}, as shown at the top of this page.
Applying eigenvalue decomposition, we can write $\mathbf{S}_n^H\boldsymbol\Sigma_{\boldsymbol\alpha}^{(1)-1}
\mathbf{S}_n$ as $\mathbf{U}_n{\rm diag}(\mathbf{v})\mathbf{U}_n^H$, where $\mathbf{v} \triangleq (v_p)_{p\in\mathcal{P}} \in \mathbb{R}^{P}$ represents the eigenvalues and $\mathbf{U}_n\in \mathbb{C}^{P\times P}$ represents the corresponding eigenvectors. For all $p\in\mathcal{P}$,
let $u_p$ denote the $p$-th diagonal element of $\mathbf{U}_n\mathbf{S}_n^H\boldsymbol\Sigma_{\boldsymbol\alpha}^{(1)-1}
\widehat{\mathbf \Sigma}_{\mathbf R}\boldsymbol\Sigma_{\boldsymbol\alpha}^{(1)-1}\mathbf{S}_n\mathbf{U}_n^H$.
Define $\mathbf{v}_{-p} \triangleq (v_{p'})_{p'\in\mathcal{P},p'\neq p} \in\mathbb{R}^{P-1}$,
\begin{align*}
\mathcal S(t) \triangleq \bigg\{(\mathbf{x},\mathbf{y})|  & \mathbf{x},\mathbf{y} \in \{0,1\}^{P-1}, x_p+y_p\leq 1,  p\in\mathcal{P}\backslash \{P\},\\ &\sum_{p\in\mathcal{P}\backslash \{P\}}(x_p+2y_p)=t \bigg\}, \\
h(\mathbf{z},t)\triangleq & \sum_{(\mathbf{x},\mathbf{y})\in \mathcal S(t)}\prod_{p=1}^{P-1}2^{x_p}z_p^{x_p+2y_p},\ \mathbf{z} \in \mathbb{R}^{P-1}_{++},
\end{align*}
where $t=0,...,2P-2$.
Based on the above definitions, we define
$g^{(1)}_{\boldsymbol\alpha,n}(d)$ in \eqref{eq:derivative_function}, as shown at the top of this page.
    \begin{Thm}[Optimal Solution of Coordinate Descent Optimization in~\eqref{eqn:Penal_a_ML_extension}]\label{Thm:Step_APs_Penal_M_non_extension}
Given $\boldsymbol\alpha$, the optimal solution of the problem in~\eqref{eqn:Penal_a_ML_extension} is given by:
\setcounter{equation}{12}
\begin{align}
d_n^{(1)*} \triangleq \arg\min\limits_{d\in\mathcal{D}^{(1)}_n\cup \{-\alpha_n,1-\alpha_n\}} f^{(1)}_{\boldsymbol\alpha,n}(d),\label{eqn:d_Penalty_a_extension}
\end{align}
where $\mathcal{D}^{(1)}_n\triangleq \{d\in[-\alpha_n,1-\alpha_n]:g^{(1)}_{\boldsymbol\alpha,n}(d) = 0\}$.
\end{Thm}
\addtolength{\topmargin}{0.01in}
\begin{IEEEproof}[Proof (Sketch)]
First, by~\eqref{eq:sigma_alpha_1}, \eqref{eqn:f_ml}, and $(\boldsymbol \Sigma^{(1)}_{\boldsymbol\alpha}+dg_{n}\mathbf S_n\mathbf S_n^H)^{-1}=\mathbf \Sigma_{\boldsymbol\alpha}^{(1)-1}-dg_n\boldsymbol\Sigma_{\boldsymbol\alpha}^{(1)-1}\mathbf{S}_n(\mathbf{I}_P+dg_n\mathbf{S}_n^H\boldsymbol\Sigma_{\boldsymbol\alpha}^{(1)-1}
\mathbf{S}_n)^{-1}\mathbf{S}_n^H\boldsymbol\Sigma_{\boldsymbol\alpha}^{-1}
$ \cite{Caire18ISIT}, we show $f^{(1)}(\boldsymbol\alpha+de_n)=
f^{(1)}(\boldsymbol\alpha)+f^{(1)}_{\boldsymbol\alpha,n}(d)$.
Thus, the problem in~\eqref{eqn:Penal_a_ML_extension} is equivalent to $\min\limits_{d\in\mathcal{D}^{(1)}_n\cup \{-\alpha_n,1-\alpha_n\}}f^{(1)}_{\boldsymbol\alpha,n}(d)$.
Next, based on eigenvalue decomposition, we show
$\left(f^{(1)}_{\boldsymbol\alpha,n}(d)\right)^{'} =
\frac{g^{(1)}_{\boldsymbol\alpha,n}(d)}{\prod_{p\in\mathcal{P}}
  (1+v_pd)^2}$. Thus, the optimal solution of $\min\limits_{d\in\mathcal{D}^{(1)}_n\cup \{-\alpha_n,1-\alpha_n\}}f^{(1)}_{\boldsymbol\alpha,n}(d)$ is given by \eqref{eqn:d_Penalty_a_extension}. Therefore, we complete the proof.
\end{IEEEproof}

$g^{(1)}_{\boldsymbol\alpha,n}(d)$ is a polynomial with degree $2P-1$ and hence has $2P-1$ roots. Note that the roots of a polynomial with degree $q$ can be obtained analytically if $q\in\{1,2,3,4\}$ and numerically otherwise\blue{\cite{press2007numerical}}. Besides, note that the computational complexities for obtaining the roots of a polynomial with degree $q$ analytically and numerically are $\mathcal{O}(q)$ and $\mathcal{O}(q^3)$, respectively \blue{\cite{press2007numerical}}. Thus, $\mathcal{D}^{(1)}_n$ can be obtained in closed-form with computational complexity $\mathcal{O}(P)$ if $P\in\{1,2\}$ and numerically with computational complexity $\mathcal{O}(P^3)$ otherwise.
The details of the coordinate descent algorithm are summarized in Algorithm~\ref{alg:ML_descend_non_extension}.
\begin{algorithm}[t] \caption{Coordinate Descent Algorithm for Problem~\ref{Prob:ML_a_non_extension}}
\hspace*{0.02in} \blue{{\bf Input:}
empirical covariance matrix $\widehat{\mathbf \Sigma}_{\mathbf R}$.} \\ \hspace*{0.02in} \blue{{\bf Output:} $\boldsymbol\alpha$.}
\begin{algorithmic}[1]
\STATE Initialize $\mathbf \Sigma^{(1)-1}_{\boldsymbol\alpha}=\frac{1}{\sigma^2} \mathbf I_L$, $\boldsymbol\alpha=\mathbf 0$.
\STATE \textbf{repeat}
\FOR {$n\in\mathcal{N}$}
\STATE Calculate $d^{(1)*}_n$ according to \eqref{eqn:d_Penalty_a_extension} analytically if $P\leq 2$ and numerically if $P\geq 3$.
\STATE \textbf{If} $d^{(1)*}_n \neq 0$
\STATE \quad Update $\alpha_{n}=\alpha_{n}+d^{(1)*}_n$.
\STATE \quad Update $\boldsymbol\Sigma^{(1)-1}_{\boldsymbol\alpha} = \boldsymbol \Sigma^{(1)-1}_{\boldsymbol\alpha}-d^{(1)*}_ng_n
\boldsymbol \Sigma_{\boldsymbol\alpha}^{(1)-1}\mathbf S_{n}
(\mathbf{I}_P+ \quad d^{(1)*}_ng_n\mathbf S_{n}^H\boldsymbol\Sigma^{(1)-1}_{\boldsymbol\alpha}\mathbf S_{n})^{-1}\mathbf S_{n}^H\boldsymbol\Sigma^{(1)-1}_{\boldsymbol\alpha}$.
\STATE \textbf{end}
\ENDFOR
\STATE \textbf{until} $\boldsymbol\alpha$ satisfies some stopping criterion.
\end{algorithmic}\label{alg:ML_descend_non_extension}
\end{algorithm}
If each coordinate optimization in \eqref{eqn:Penal_a_ML_extension} has a unique optimal solution, Algorithm~\ref{alg:ML_descend_non_extension} converges to a stationary point of Problem~\ref{Prob:ML_a_non_extension}, as the number of the iteration goes to infinity \cite[Proposition 2.7.1]{Bertsekas99}.
The complexities of Step 4, Step 6, and Step 7 are $\mathcal{O}(PL^2)$, $\mathcal{O}(1)$, and $\mathcal{O}(PL^2)$, respectively (note that $P<L$).
Thus, the computational complexity of each iteration of Algorithm~\ref{alg:ML_descend_non_extension} is $\mathcal{O}(NPL^2)$.
\setcounter{TempEqCnt}{\value{equation}}
\setcounter{equation}{22}
\begin{figure*}
\begin{align}
&f^{(2)}_{\mathbf{b},i}(d) \triangleq \log(1+dg_i\mathbf S_{:,i}^H\mathbf \Sigma^{-1}_{\boldsymbol\beta}\mathbf S_{:,i}) -\frac{dg_i\mathbf S_{:,i}^H\mathbf \Sigma^{-1}_{\boldsymbol\beta}\widehat{\mathbf \Sigma}_{\mathbf R}\mathbf \Sigma^{-1}_{\boldsymbol\beta}\mathbf S_{:,i} }{1+dg_i\mathbf S_{:,i}^H\mathbf \Sigma^{-1}_{\boldsymbol\beta}\mathbf S_{:,i}}+\frac{\rho d}{P}
\bigg(1-\frac{d}{P}-\frac{2}{P}\sum\limits_{p=1}^{P}\beta_{\left(\lceil \frac{i}{P} \rceil-1\right)P+p}\bigg) \label{eq:f_old}\\
 & A_i\triangleq -\frac{2\rho g_i^2}{P^2}\left(\mathbf S_{:,i}^H\mathbf \Sigma^{-1}_{\boldsymbol\beta}\mathbf S_{:,i}\right)^2, \quad
B_i\triangleq \frac{\rho g_i^2}{P}\bigg(1-\frac{2}{P}\sum\limits_{p=1}^{P}\beta_{\left(\lceil \frac{i}{P} \rceil-1\right)P+p}\bigg)\left(\mathbf S_{:,i}^H\mathbf \Sigma^{-1}_{\boldsymbol\beta}\mathbf S_{:,i}\right)^2-\frac{4\rho g_i}{P^2}\mathbf S_{:,i}^H\mathbf \Sigma^{-1}_{\boldsymbol\beta}\mathbf S_{:,i}\notag \\
&C_i\triangleq g_i^2 \left(\mathbf S_{:,i}^H\mathbf \Sigma^{-1}_{\boldsymbol\beta}\mathbf S_{:,i}\right)^2+\frac{2\rho g_i}{P}\bigg(1-\frac{2}{P}\sum\limits_{p=1}^{P}\beta_{\left(\lceil \frac{i}{P} \rceil-1\right)P+p}\bigg)\mathbf S_{:,i}^H\mathbf \Sigma^{-1}_{\boldsymbol\beta}\mathbf S_{:,i}-\frac{2\rho}{P^2}\notag \\
 &D_i\triangleq g_i\mathbf S_{:,i}^H\mathbf \Sigma^{-1}_{\boldsymbol\beta}\mathbf S_{:,i} -g_i\mathbf S_{:,i}^H\mathbf \Sigma^{-1}_{\boldsymbol\beta}\widehat{\mathbf \Sigma}_{\mathbf R}\mathbf \Sigma^{-1}_{\boldsymbol\beta}\mathbf S_{:,i} +\frac{\rho}{P}\bigg(1-\frac{2}{P}\sum\limits_{p=1}^{P}\beta_{\left(\lceil \frac{i}{P} \rceil-1\right)P+p}\bigg)\notag
\end{align}
\hrulefill
\vspace{-1mm}
\end{figure*}
\setcounter{equation}{\value{TempEqCnt}}
\section{MLE-based Device Activity Detection Using Penalty Method and Coordinate Descent Method}\label{sec:ML}
In this section, we propose an MLE-based device activity detection method based on an equivalent \blue{form} of the received signal $\mathbf{r}_m$ in \eqref{eqn:received_signal_time}. Notice that based on this equivalent \blue{form}, conventional channel estimation methods can be directly applied for channel estimation of detected active devices under frequency-selective Rayleigh fading.
\blue{\subsection{Problem Formulation}
First, we formulate an MLE problem for activity detection of $NP$ virtual devices. Let $\mathcal{I} \triangleq \{1,...,NP\}$ denote the set of virtual devices. Let $\beta_i$ denote the activity states of virtual device $i$, for all $i\in\mathcal{I}$. Virtual devices $(n-1)P+1,...,nP$ share the same activity state and channel condition as actual device $n$, for all $n\in\mathcal{N}$. Thus, we have:
\begin{align}
 & \beta_{(n-1)P+1}=...=\beta_{nP}, \  n\in\mathcal{N},\label{eq:equality_constraints} \\
 &   \beta_i \in [0,1], i\in\mathcal{I}, \label{eqn_beta_n}\\
& \alpha_n = \frac{\sum_{p\in\mathcal{P}}\beta_{(n-1)P+p}}{P},\quad n\in\mathcal{N}. \label{eq:construct_alpha}
\end{align}
Therefore, the received signal $\mathbf{r}_m$ from the $N$ devices, $\mathbf{r}_m$ in \eqref{eqn:received_signal_time}, can be equivalently rewritten as the received signal from the $NP$ virtual devices as follows:}
%
\begin{align}
\mathbf r_m
= & \mathbf S\mathbf B \mathbf{G}^{\frac{1}{2}}\mathbf h_m+\mathbf n_m, \quad m\in\mathcal{M}, \label{eqn:received_signal_time_xx}
\end{align}

where
$\mathbf{S}\triangleq \left[ \mathbf{S}_1,...,\mathbf{S}_N \right]$ $\in\mathbb{C}^{L\times NP}$,
$\mathbf B\triangleq {\rm diag}\left(\boldsymbol\beta\right)$ with $\boldsymbol\beta \triangleq (\beta_i)_{i\in\mathcal{I}}$, $\mathbf{G}\triangleq \rm{diag}\left(\mathbf{g}
\right)\otimes\mathbf I_P \in \mathbb{R}_{++}^{NP\times NP}$ with $\mathbf{g}\triangleq (g_n)_{n\in\mathcal{N}} \in \mathbb{R}_+^{N}$. 
Noting that $\mathbf h_m \triangleq \left[\mathbf h^T_{1,m},...,\mathbf h^T_{N,m}\right]^T \in \mathbb{C}^{NP}$,
$\mathbf h_m$, $m\in\mathcal M$ are i.i.d. according to ${\mathcal CN}(\mathbf 0,\mathbf I_{NP})$. Thus, when  $\beta_i,i\in\mathcal{I}$, $g_n,n\in\mathcal{N}$ are given, $\mathbf r_m, m \in\mathcal{M}$, with $\mathbf{r}_m$ given by \eqref{eqn:received_signal_time_xx}, are i.i.d. according to $ \mathcal{CN}\left(\mathbf 0,\boldsymbol\Sigma^{(2)}_{\boldsymbol\beta} \right)$ \cite{Caire18ISIT}, where
\begin{align}
\boldsymbol\Sigma^{(2)}_{\boldsymbol\beta} \triangleq \mathbf S\mathbf B\mathbf{G}\mathbf S^H+\sigma^2\mathbf I_L. \label{eq:sigma_alpha_2}
\end{align}
Thus, the likelihood function of $\mathbf R$, viewed as a function of $\boldsymbol\beta$, can also be expressed as:
\begin{align}
&p^{(2)}(\mathbf R;\boldsymbol\beta) \triangleq \frac{\exp\left(-\text{tr}\left(\boldsymbol\Sigma_{\boldsymbol\beta}^{(2)-1}\mathbf R\mathbf R^{H}\right)\right)}{\pi^{LM}\vert\boldsymbol\Sigma
^{(2)}_{\boldsymbol\beta}\vert^M}.\label{eqn:likelihood_no_coop}
\end{align}
The maximization of $p^{(2)}(\mathbf R;\boldsymbol\beta)$ is equivalent to the minimization of $f^{(2)}(\boldsymbol\beta)$, where
\begin{align}
 f^{(2)}(\boldsymbol\beta) \triangleq & -\log p^{(2)}(\mathbf R;\boldsymbol\beta)-L\log\pi \nonumber \\ =&\log|\boldsymbol\Sigma^{(2)}_{\boldsymbol\beta}|+\text{tr}\left(\boldsymbol\Sigma_{\boldsymbol\beta}^{(2)-1}\widehat{\mathbf \Sigma}_{\mathbf R}\right). \label{eqn:f_ml_ex}
\end{align}
Thus, the MLE problem of $\boldsymbol\beta$ from $\mathbf{R}$ given by \eqref{eqn:likelihood_no_coop} can be formulated as follows.
\begin{Prob}[MLE for Activity Detection of Virtual Devices]\label{Prob:ML_a}
\begin{align}
\min_{\boldsymbol\beta} &\quad f^{(2)}(\boldsymbol\beta)\notag\\
s.t. &\quad   \eqref{eq:equality_constraints}, \quad \eqref{eqn_beta_n}.
\end{align}
\end{Prob}
Problem~\ref{Prob:ML_a} is also a non-convex optimization problem. It differentiates from Problem~\ref{Prob:ML_a_non_extension}, as $\boldsymbol\Sigma^{(1)}_{\boldsymbol\alpha}$ and $\boldsymbol\Sigma^{(2)}_{\boldsymbol\beta}$ have different forms. Besides, the objective function of Problem~\ref{Prob:ML_a} shares the same form as the objective function of the MLE problem for activity detection of $NP$ devices under flat Rayleigh fading in \cite{Caire18ISIT} except that the dimensions of $\mathbf{S}\in\mathbb{C}^{L\times NP}$, $\mathbf{B}\in\mathbb{C}^{NP \times NP}$, $\mathbf{G}\in\mathbb{C}^{NP\times NP}$, and
$\mathbf{h}_m\in\mathbb{C}^{NP}$. However, unlike Problem~\ref{Prob:ML_a_non_extension} and the ML estimation problem in \cite{Caire18ISIT}, Problem~\ref{Prob:ML_a} has  extra coupling constraints in  \eqref{eq:equality_constraints}. To address the issue caused by the coupling constraints in \eqref{eq:equality_constraints}, we apply the penalty method \cite{Bertsekas99} to obtain a stationary point of an equivalent problem of Problem~\ref{Prob:ML_a} in Section~\ref{subsec:penal}.
Later in Section~\ref{sec:simulation}, we shall see that the device activity detection method based on the penalty method has a higher accuracy and
a higher computational complexity than the method based on relaxation. After solving Problem~\ref{Prob:ML_a} for $\boldsymbol\beta$, we can construct a device activities of the $N$ actual devices $\boldsymbol\alpha$ according to \eqref{eq:construct_alpha}
\subsection{Solution}\label{subsec:penal}
We disregard the coupling constraints in \eqref{eq:equality_constraints} and add to the objective function of Problem~\ref{Prob:ML_a} a penalty for violating them. Then, we can convert Problem~\ref{Prob:ML_a} to the following problem.
\begin{Prob}[Penalty Problem of Problem~\ref{Prob:ML_a}]\label{Prob:penal}
\begin{align*}
\min_{\mathbf b}\quad & \tilde{f}^{(2)}(\boldsymbol\beta)\triangleq f^{(2)}(\boldsymbol\beta)
+\rho \eta(\boldsymbol\beta),\notag \\
s.t. \quad &  \eqref{eqn_beta_n},
\end{align*}
where $\rho>0$ is the penalty parameter, and
\begin{align}
\eta(\boldsymbol\beta) \triangleq \sum_{n\in\mathcal{N}} \frac{\sum\limits_{p\in\mathcal{P}}\beta_{(n-1)P+p}}{P}
\bigg(1-\frac{\sum\limits_{p\in\mathcal{P}}\beta_{(n-1)P+p}}{P}\bigg) \label{eq:penaltyfunction}
\end{align}
is the penalty function.
\end{Prob}

If $\rho$ is sufficiently large, an optimal solution of Problem~\ref{Prob:penal} is also optimal for Problem~\ref{Prob:ML_a} (as $f^{(2)}(\boldsymbol\beta)$ is bounded from above) \cite{Bertsekas99}.
Now, we adopt the coordinate descent method to obtain a stationary point of Problem~\ref{Prob:penal} instead of Problem~\ref{Prob:ML_a}. Specifically, given $\boldsymbol\beta$ obtained in the previous step, the coordinate descent optimization with respect to $\beta_i$ is equivalent to the optimization of the increment $d$ in $\beta_i$:
\begin{align}
\min_{d\in [-\beta_i,1-\beta_i]} \ \tilde{f}^{(2)}(\boldsymbol\beta + d\mathbf e_i).\label{eqn:Penal_a}
\end{align}
We shall see that it is more challenging to solve the coordinate optimization in \eqref{eq:three_times} than to solve the MLE problem for flat Rayleigh fading in \cite{Caire18ISIT}. Similarly, we define two important functions before solving the problem \eqref{eqn:Penal_a}, i.e., $f^{(2)}_{\boldsymbol\beta,i}(d)$ in \eqref{eq:f_old}, as shown at the top of this  page, and
\setcounter{equation}{23}
\begin{align}\label{eq:three_times}
  \tilde{g}^{(2)}_{\boldsymbol\beta,i}(d) \triangleq
   A_id^3+B_id^2+C_id+D_i,
\end{align}
where $A_i,B_i,C_i,D_i$ are given at the top of this  page. \blue{Note that $\tilde{g}^{(2)}_{\boldsymbol\beta,i}(d)$ is the numerator of the derivative of $\tilde{f}^{(2)}_{\boldsymbol\beta,i}(d)$ (which is a fraction). By taking  the derivative of $\tilde{f}^{(2)}_{\boldsymbol\beta,i}(d)$, simplifying it based on activity detection, and setting the simplified derivative of $\tilde{f}^{(2)}_{\boldsymbol\beta,i}(d)$ to zero, we derive
the optimal solution of the problem in~\eqref{eqn:Penal_a}, which is expressed in terms of $\tilde{f}^{(2)}_{\boldsymbol\beta,i}(d)$ and $\tilde{g}^{(2)}_{\boldsymbol\beta,i}(d)$.}
\begin{Thm}[Optimal Solution of Coordinate Descent Optimizations in~\eqref{eqn:Penal_a}]\label{Thm:Step_APs_Penal_M}
Given $\boldsymbol\beta$, the optimal solution of the problem in~\eqref{eqn:Penal_a} is given by:
\begin{align}
d^{(2)*}_i = \arg\min\limits_{d\in\mathcal{D}^{(2)}_i\cup\{-\beta_i,1-\beta_i\}} f^{(2)}_{\boldsymbol\beta,i}(d),\label{eqn:d_Penalty_a}
\end{align}
where $\mathcal{D}^{(2)}_i \triangleq \{d\in[-\beta_i,1-\beta_i]:g^{(2)}_{\boldsymbol\beta,i}(d) = 0\}$.
\end{Thm}
\begin{IEEEproof}[Proof (Sketch)]
By \eqref{eq:sigma_alpha_2}, \eqref{eqn:f_ml_ex}, and $(\mathbf \Sigma_{\boldsymbol\beta}+dg_{i}\mathbf S_{:,i}\mathbf S_{i,:}^H)^{-1}=\mathbf \Sigma^{-1}_{\boldsymbol\beta}-\frac{dg_i\mathbf \Sigma^{-1}_{\boldsymbol\beta}\mathbf{S}_{:,i}\mathbf{S}_{i,:}^H\mathbf \Sigma^{-1}_{\boldsymbol\beta}}{1+dg_i\mathbf{S}_{i,:}^H\mathbf \Sigma^{-1}_{\boldsymbol\beta}\mathbf{S}_{:,i}}$ \cite{Caire18ISIT}, we show $\tilde{f}^{(2)}(\boldsymbol\beta+de_i)=
\tilde{f}^{(2)}(\boldsymbol\beta)+f^{(2)}_{\boldsymbol\beta,i}(d)$.
Thus, the problem in~\eqref{eqn:Penal_a} is equivalent to $\min\limits_{d\in\mathcal{D}^{(2)}_i\cup \{-\beta_i,1-\beta_i\}}f^{(2)}_{\boldsymbol\beta,i}(d)$.
Next, following the derivation of (22) in \cite{Caire18ISIT}, we show
$\left(f^{(2)}_{\boldsymbol\beta,i}(d)\right)^{'} =
  \frac{g_{\boldsymbol\beta,i}(d)}{(1+dg_i\mathbf{S}_{i,:}^H\mathbf \Sigma^{-1}_{\boldsymbol\beta}\mathbf{S}_{:,i})^2}$. Thus, the optimal solution of $\min\limits_{d\in\mathcal{D}^{(2)}_i\cup \{-\beta_i,1-\beta_i\}}f^{(2)}_{\boldsymbol\beta,i}(d)$ is given by \eqref{eqn:d_Penalty_a}. Therefore, we complete the proof.
\end{IEEEproof}

As $g^{(2)}_{\boldsymbol\beta,i}(d)$ is a polynomial with degree 3, $\mathcal{D}^{(2)}_i$ can be obtained in closed-form with computational complexity $\mathcal{O}(P)$.

The details of the coordinate descent algorithm are summarized in Algorithm~\ref{alg:ML_descend}.
\begin{algorithm}[t] \caption{Coordinate Descent Algorithm for Problem~\ref{Prob:penal}}
\hspace*{0.02in} \blue{{\bf Input:}
empirical covariance matrix $\widehat{\mathbf \Sigma}_{\mathbf R}$.} \\ \hspace*{0.02in} \blue{{\bf Output:} $\boldsymbol\beta$.}
\begin{algorithmic}[1]
\STATE Initialize $\mathbf \Sigma^{-1}_{\boldsymbol\beta}=\frac{1}{\sigma^2} \mathbf I_L$, $\boldsymbol\beta=\mathbf 0$.
\STATE \textbf{repeat}
\FOR {$i\in\mathcal{I}$}
\STATE Calculate $d^{(2)*}_i$ according to \eqref{eqn:d_Penalty_a}.
\STATE \textbf{If} $d^{(2)*}_i\neq 0$
\STATE \quad Update $\beta_{i}=\beta_{i}+d^{(2)*}_i$.
\STATE \quad Update $\mathbf \Sigma^{-1}_{\boldsymbol\beta} = \mathbf \Sigma^{-1}_{\boldsymbol\beta}-\frac{d^{(2)*}_ig_i\mathbf \Sigma_{\boldsymbol\beta}^{-1}\mathbf S_{:,i}\mathbf S_{:,i}^H\mathbf \Sigma^{-1}_{\boldsymbol\beta}}{1+d^{(2)*}_ig_i\mathbf S_{:,i}^H\mathbf \Sigma^{-1}_{\boldsymbol\beta}\mathbf S_{:,i}}$.
\STATE \textbf{end}
\ENDFOR
\STATE \textbf{until} $\boldsymbol\beta$ satisfies some stopping criterion.
\end{algorithmic}\label{alg:ML_descend}
\end{algorithm}
If each coordinate optimization in \eqref{eqn:Penal_a} has a unique optimal solution, Algorithm~\ref{alg:ML_descend} converges to a stationary point of Problem~\ref{Prob:penal} as the number of iteration goes to infinity \cite[Proposition 2.7.1]{Bertsekas99}.
The computational complexities of Step 4, Step 6, and Step 7 are $\mathcal{O}(L^2)$, $\mathcal{O}(1)$, and $\mathcal{O}(L^2)$, respectively. Thus, the computational complexity of each iteration of Algorithm \ref{alg:ML_descend} is $\mathcal{O}\left(NPL^2\right)$.
\addtolength{\topmargin}{0.025in}
\section{Numerical Results}\label{sec:simulation}
In this section, we evaluate the performance of the proposed MLE-based device activity detection methods given by Algorithm~\ref{alg:ML_descend_non_extension} and Algorithm~\ref{alg:ML_descend}, referred to as {\em Prop.-MLE-Alg. 1} and {\em Prop.-MLE-Alg. 2}, respectively. We consider three baseline schemes, namely, {\em BL-MLE}, {\em BL-GL}, and {\em BL-AMP}, which are obtained by applying the existing MLE \cite{Yu19ICC}, GROUP LASSO \cite{JSAC_li}, and AMP \cite{Liu18TSP}, proposed for flat Rayleigh fading, to detect the activities of the $NP$ virtual devices, $\boldsymbol\beta$, without considering the constraints in \eqref{eq:equality_constraints}, and then setting the activities of the $N$ actual devices, $\boldsymbol\alpha$, according to $\alpha_n = \frac{\sum\limits_{p\in\mathcal{P}}\beta_{(n-1)P+p}}{P},n\in\mathcal{N}$. The thresholds for the MLE-based schemes and {\em BL-GL} are numerically optimized. The threshold for {\em BL-AMP} is chosen as in \cite{Liu18TSP}. We generate pilots according to i.i.d. $\mathcal{CN}(0,\mathbf{I}_L)$ and normalize their norms to $\sqrt{L}$ \cite{Caire18ISIT,Yu19ICC}.
In the simulation, we independently generate $1000$ realizations for $\alpha_n \sim \text{B}(1000,0.07)$, $n\in\mathcal N$, $ h_{n,m,p} \sim\mathcal{CN}(0,1)$, $n\in\mathcal N$, $m\in\mathcal M$, $p\in\mathcal P$, and Gaussian pilots in each realization and evaluate the average error rate over all $1000$ realizations.
Unless otherwise stated, we choose $N=1000$, $L=72$, $M=128$, $P=4$, $g_n=1,n\in\mathcal{N}$, and $\sigma^2=0.1$.

\begin{figure}[t]
\begin{center}
\subfigure[\scriptsize{Number of channel taps $P$.
}\label{fig:error_vs_P}]
{\resizebox{4.252cm}{!}{\includegraphics{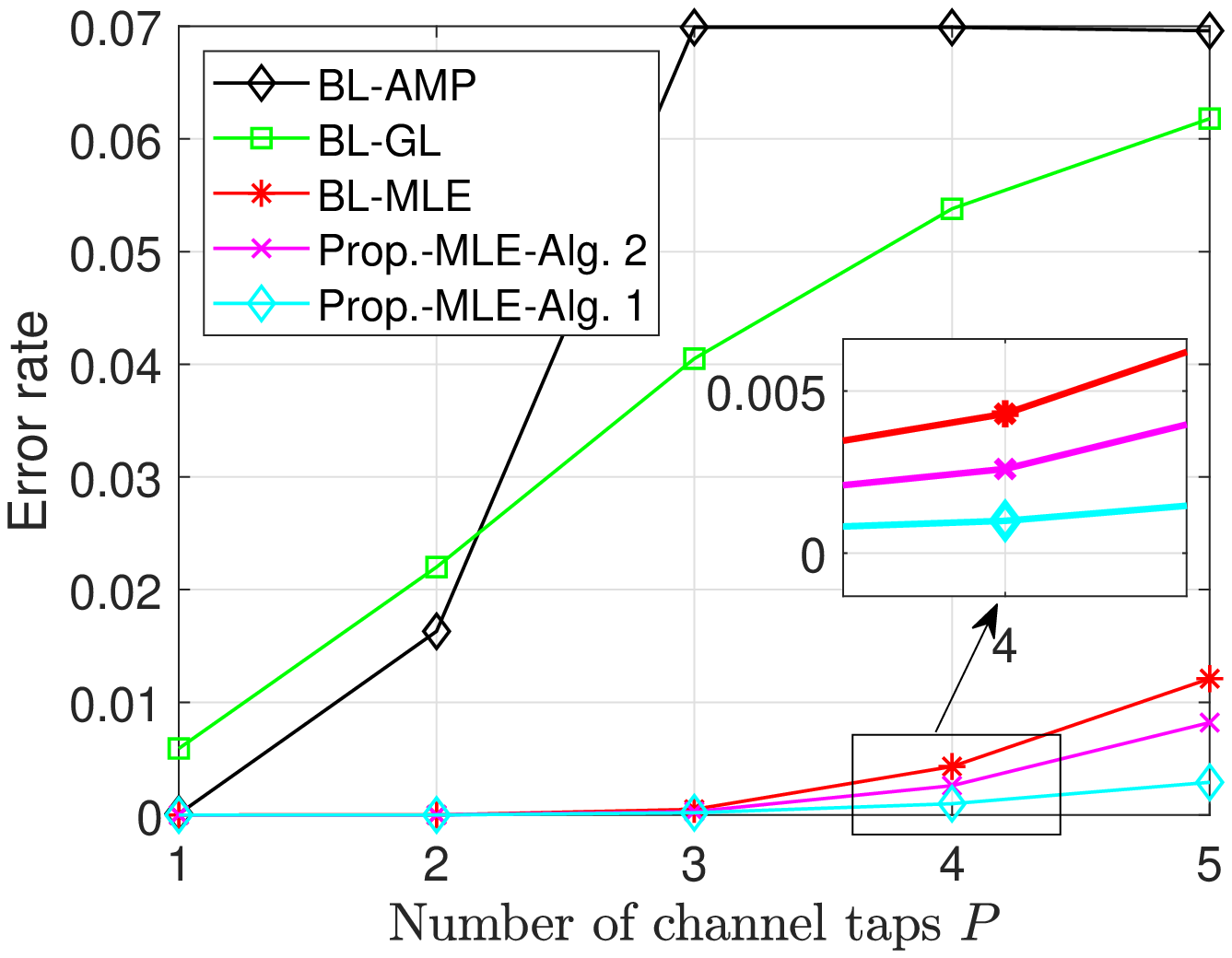}}}\quad
\subfigure[\scriptsize{Pilot length $L$.
}\label{fig:error_vs_L}]
{\resizebox{4.252cm}{!}{\includegraphics{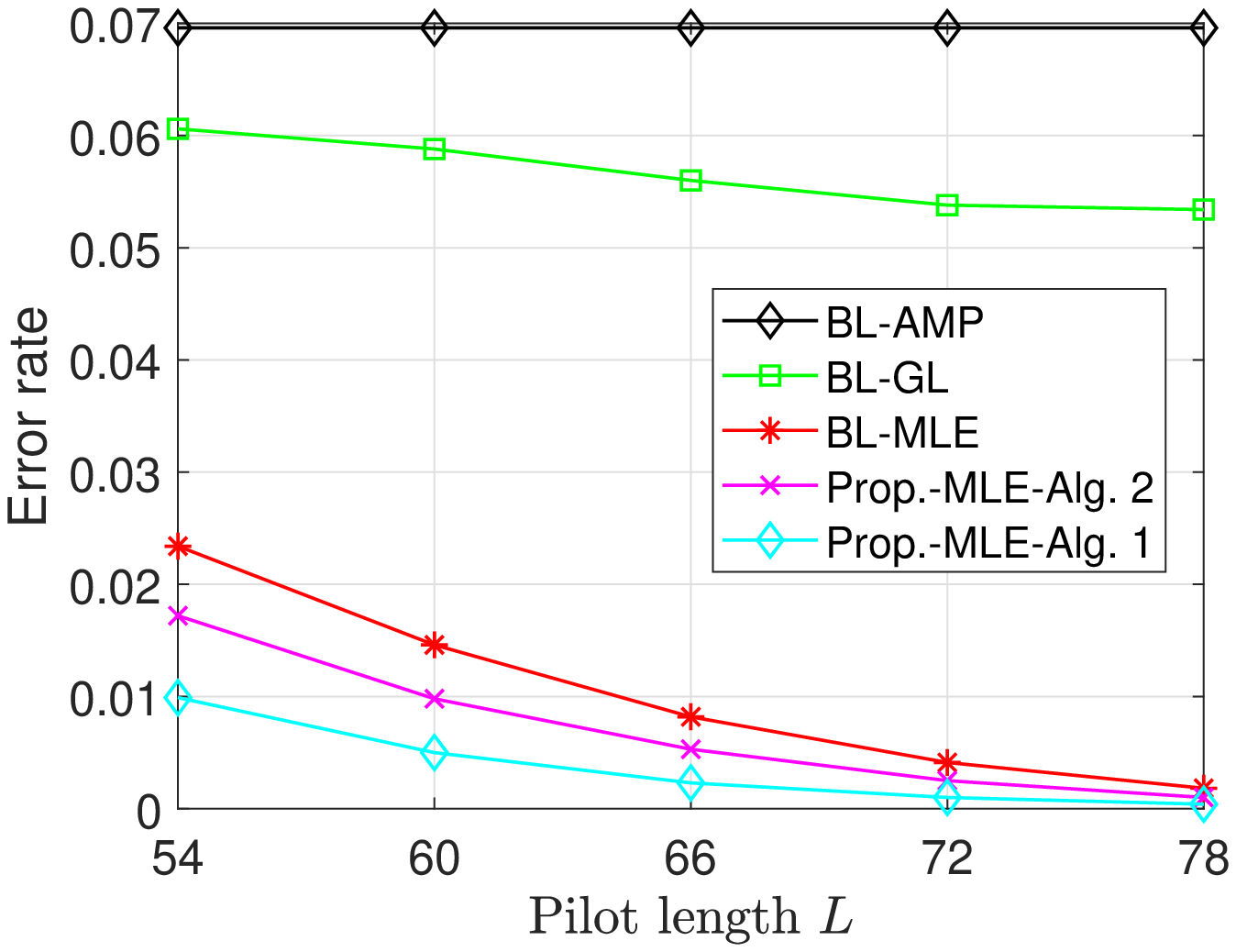}}}
\end{center}

\caption{\small{Error rate versus number of channel taps $P$ and pilot length $L$.}}
\vspace{-4mm}
\label{fig:rate}
\end{figure}

Fig.~\ref{fig:error_vs_P} and Fig.~\ref{fig:error_vs_L} plot the error rate versus the number of channel taps $P$ and the pilot length $L$, respectively.
From the two figures, we can make the following observations. The MLE-based schemes significantly outperform the compressed sensing-based schemes, {\em BL-AMP} and {\em BL-GL}. Note that at small $L$, {\em BL-AMP} does not work properly, yielding a poor error rate. The two proposed MLE-based schemes outperform {\em BL-MLE}, as they rigorously tackle the MLE problems. {\em Prop.-MLE-Alg. 1} for solving Problem~\ref{Prob:ML_a_non_extension} with size $N$ achieves a smaller error rate than {\em Prop.-MLE-Alg. 2} for solving Problem~\ref{Prob:penal} with size $NP$, as a problem with a smaller size can be more effectively solved. Besides, the error rates of most schemes increase with $P$ and decrease with $L$. The slight increase of the error rate of {\em Prop.-MLE-Alg. 1} with $P$ when $P\geq 3$ is mainly due to the numerical error for determining $\mathcal{D}_n^{(1)}, n\in\mathcal{N}$. The increase of the error rates of the  other schemes with $P$ derives from the increase in the number of virtual devices.

\begin{figure}[t]
\begin{center}
\subfigure[\small{Length of channel taps $P$ at $L=64$ and $M=128$.}\label{fig:error_vs_time1}]
{\resizebox{4.252cm}{!}{\includegraphics{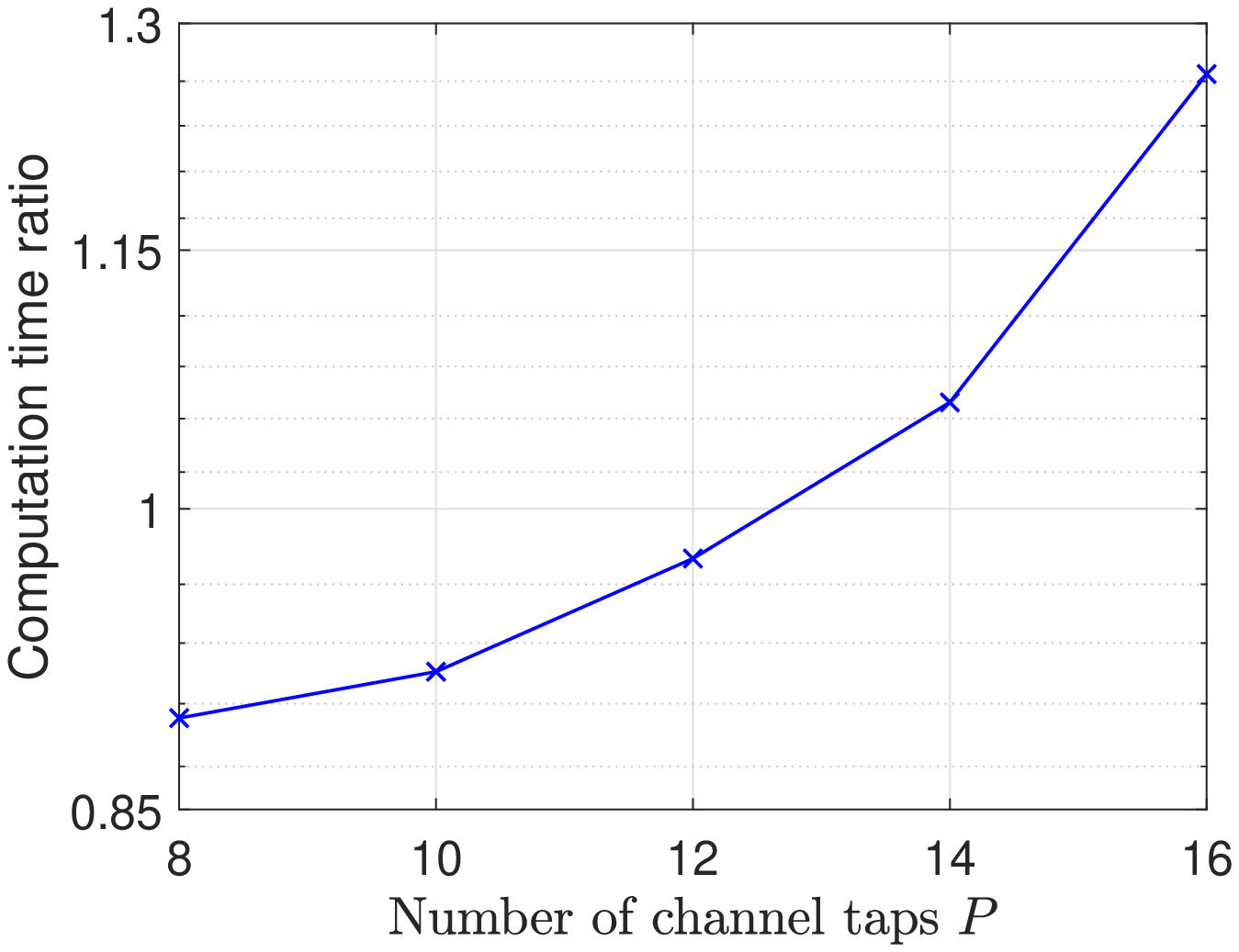}}}\quad
\subfigure[\small{Length of channel taps $P$ at $L=32$ and $M=256$.}\label{fig:error_vs_time2}]
{\resizebox{4.252cm}{!}{\includegraphics{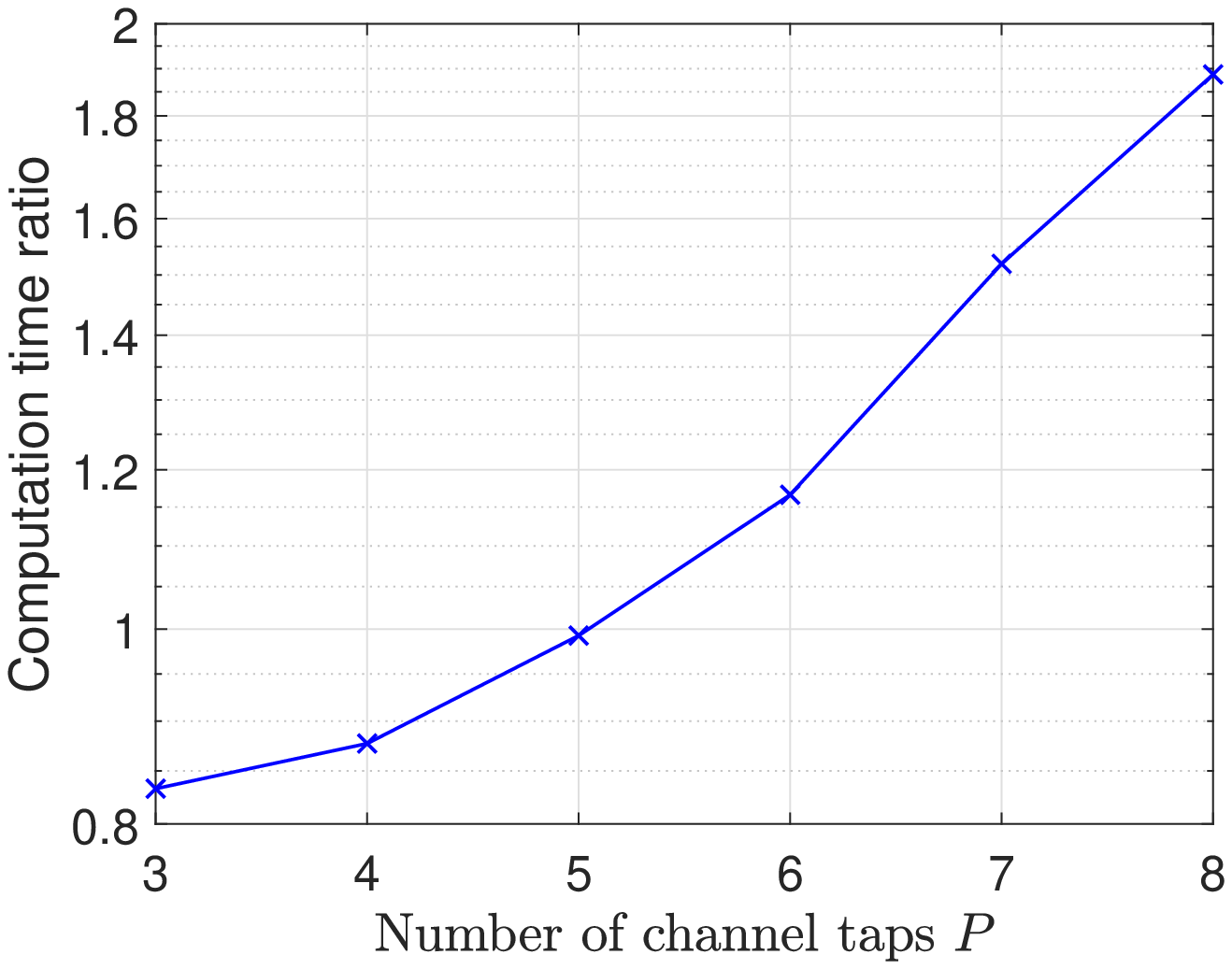}}}
\end{center}
\vspace{-2mm}
\caption{\small{Computation time ratio versus number of channel taps $P$.}}
\vspace{-4mm}
\label{fig:gamma_approx_coop}
\end{figure}

Fig.~\ref{fig:error_vs_time1} and Fig.~\ref{fig:error_vs_time2} plot the ratio between the computation time of {\em Prop.-MLE-Alg. 1} and the computation time of {\em Prop.-MLE-Alg. 2} versus the number of channel taps $P$ at different pilot lengths.
{\em Prop.-MLE-Alg. 1} has shorter computation time than {\em Prop.-MLE-Alg. 2} at small $P$, as the overall computation time for determining $\mathcal{D}_n^{(1)},n\in\mathcal{N}$ analytically is short at small $P$. {\em Prop.-MLE-Alg. 1} has larger computation time than {\em Prop.-MLE-Alg. 2} at large $P$, as the overall computation time for determining $\mathcal{D}_i^{(2)},i\in\mathcal{I}$ analytically is shorter at large $P$. When $L$ is large, {\em Prop.-MLE-Alg. 1} outperforms {\em Prop.-MLE-Alg. 2} at most practical values of $P$.


\section{Conclusion}
In this paper, we first presented an OFDM-based massive grant-free access scheme for a wideband system. Then, we proposed two MLE-based device activity detection methods for frequency-selective Rayleigh fading using statistical estimation and optimization techniques. The two proposed methods have different preferable system parameters and include the existing MLE-based method for flat Rayleigh fading as a special case. Conventional channel estimation methods can be directly applied for channel estimation of detected active devices under frequency selective Rayleigh fading, based on a received pilot signal model derived in this paper.

%
%

\bibliographystyle{IEEEtran}
\bibliography{IEEEabrv,Globecom}

\begin{thebibliography}{10}
\providecommand{\url}[1]{#1}
\csname url@samestyle\endcsname
\providecommand{\newblock}{\relax}
\providecommand{\bibinfo}[2]{#2}
\providecommand{\BIBentrySTDinterwordspacing}{\spaceskip=0pt\relax}
\providecommand{\BIBentryALTinterwordstretchfactor}{4}
\providecommand{\BIBentryALTinterwordspacing}{\spaceskip=\fontdimen2\font plus
\BIBentryALTinterwordstretchfactor\fontdimen3\font minus
  \fontdimen4\font\relax}
\providecommand{\BIBforeignlanguage}[2]{{%
\expandafter\ifx\csname l@#1\endcsname\relax
\typeout{** WARNING: IEEEtran.bst: No hyphenation pattern has been}%
\typeout{** loaded for the language `#1'. Using the pattern for}%
\typeout{** the default language instead.}%
\else
\language=\csname l@#1\endcsname
\fi
#2}}
\providecommand{\BIBdecl}{\relax}
\BIBdecl

\bibitem{Erik}
L.~Liu, E.~G. Larsson, W.~Yu, P.~Popovski, C.~Stefanovic, and E.~de~Carvalho,
  ``{Sparse} {Signal} {Processing} for {Grant-}{Free} {Massive} {Connectivity}:
  {A} {Future} {Paradigm} for {Random} {Access} {Protocols} in the {Internet}
  of {Things},'' \emph{IEEE Signal Process. Mag.}, vol.~35, no.~5, pp. 88--99,
  Sept. 2018.

\bibitem{Liu18TSP}
L.~{Liu} and W.~{Yu}, ``{Massive} {Connectivity} {With} {Massive} {MIMO-}{Part}
  {I}: Device {Activity} {Detection} and {Channel} {Estimation},'' \emph{IEEE
  Trans. Signal Process.}, vol.~66, no.~11, pp. 2933--2946, Jun. 2018.

\bibitem{grouplasso}
K.~S. Z.~Qin and D.~Goldfarb, ``{Efficient} block-coordinate descent algorithms
  for the group lasso,'' \emph{Math. Program. Comput.}, vol.~5, no.~2, pp.
  340--354, Jun. 2013.

\bibitem{JSAC_li}
Y.~Cui, S.~Li, and W.~Zhang, ``{Jointly} {Sparse} {Signal} {Recovery} and
  {Support} {Recovery} via {Deep} {Learning} {With} {Applications} in
  {MIMO}-{Based} {Grant}-{Free} {Random} {Access},'' \emph{IEEE J. Sel. Areas
  Commun.}, vol.~39, no.~3, pp. 788--803, Mar. 2021.

\bibitem{Caire18ISIT}
A.~{Fengler}, S.~{Haghighatshoar}, P.~{Jung}, and G.~{Caire},
  ``{Non}-{Bayesian} {Activity} {Detection}, {Large}-{Scale} {Fading}
  {Coefficient} {Estimation}, and {Unsourced} {Random} {Access} {With} a
  {Massive} {MIMO} {Receiver},'' \emph{IEEE Trans. Inf. Theory}, vol.~67,
  no.~5, pp. 2925--2951, May 2021.

\bibitem{Yu19ICC}
Z.~{Chen}, F.~{Sohrabi}, Y.~{Liu}, and W.~{Yu}, ``{Covariance} based joint
  activity and data detection for massive random access with massive {MIMO},''
  in \emph{Proc. IEEE ICC}, May 2019, pp. 1--6.

\bibitem{Jiang21TWC}
D.~{Jiang} and Y.~{Cui}, ``{ML} and {MAP} {Device} {Activity} {Detections} for
  {Grant-}{Free} {Massive} {Access} in {Multi-}{Cell} {Networks},'' \emph{be
  submitted to IEEE TWC}, 2021.

\bibitem{8421267}
J.~{Choi}, ``On {Simultaneous} {Multipacket} {Channel} {Estimation} and
  {Reception} in {Random} {Access} for {MTC} {Under} {Frequency}-{Selective}
  {Fading},'' \emph{IEEE Trans. Commun.}, vol.~66, no.~11, pp. 5360--5369, Jul.
  2018.

\bibitem{press2007numerical}
\BIBentryALTinterwordspacing
W.~Press, W.~H, S.~Teukolsky, W.~Vetterling, S.~A, and B.~Flannery,
  \emph{Numerical Recipes 3rd Edition: The Art of Scientific Computing}.\hskip
  1em plus 0.5em minus 0.4em\relax Cambridge University Press, 2007. [Online].
  Available: \url{https://books.google.com/books?id=1aAOdzK3FegC}
\BIBentrySTDinterwordspacing

\bibitem{Bertsekas99}
D.~Bertsekas, \emph{Nonlinear Programming}.\hskip 1em plus 0.5em minus
  0.4em\relax Athena Scientific, 1999.

\end{thebibliography}

\vspace{12pt}

\end{document}